\documentclass[preprint,prl,aps,showkeys,preprintnumbers,amsmath,amssymb,nofootinbib]{revtex4}
\usepackage{graphicx}
\usepackage{bm}
\usepackage{mathrsfs}
\usepackage{amsmath}

\usepackage[active]{srcltx}
\usepackage[latin1,applemac]{inputenc}
\usepackage{natbib}
\usepackage{amsfonts}
\usepackage[english]{babel}
\usepackage{epstopdf}
\usepackage{color}
\usepackage[usenames,dvipsnames,svgnames,table]{xcolor}
\usepackage{placeins}
\usepackage{setspace}
\newcommand{\YBCO}{$\mathrm{YBa}_2\mathrm{Cu}_3\mathrm{O}_7$\;}


\begin{document}

\title{	Static and Radio-frequency magnetic response of high T$_{c}$ Superconducting Quantum Interference Filters made by ion irradiation}

\author{Eliana Recoba Pawlowski,$^{1}$ Julien Kermorvant,$^{2}$ Denis Cr\'{e}t\'{e},$^{1}$ Yves Lema\^{i}tre,$^{1}$ Bruno Marcilhac,$^{1}$ Christian Ulysse,$^{3}$ Fran\c{c}ois Cou\"edo,$^{4}$ Cheryl Feuillet-Palma,$^{4}$ Nicolas Bergeal,$^{4}$ }
\author{J\'{e}rome Lesueur,$^{4}$}
\email[Corresponding author : ]{jerome.lesueur@espci.fr}
\affiliation{$^{1}$ Unit\'e Mixte de Physique CNRS, Thales, Universit\'e Paris-Sud, Universit\'e Paris-Saclay, 91 767 Palaiseau, France.}
\affiliation{$^{2}$ Thales Communication and Security, Gennevilliers, France}
\affiliation{$^{3}$ Centre de Nanosciences et de Nanotechnologie, CNRS, Universit\'e Paris Saclay, Marcoussis, France.}
\affiliation{$^{4}$ Laboratoire de Physique et d'Etude des Mat\'eriaux, CNRS, ESPCI Paris, PSL Research University, UPMC, Paris, France.}

\begin{abstract}

Superconducting Quantum Interference Filters (SQIF) are promising devices for Radio-Frequency (RF) detection combining low noise, high sensitivity, large dynamic range and wide-band capabilities. Impressive progress have been made recently in the field, with SQIF based antennas and amplifiers showing interesting properties in the GHz range using the well-established Nb/AlOx technology. The possibility to extend these results to High Temperature Superconductors (HTS) is still open, and different techniques to fabricate HTS SQIFs are competing to make RF devices.

We report on the DC and RF response of a High Temperature SQIF fabricated by the ion irradiation technique. It is made of 1000 Superconducting QUantum Interference Devices (SQUIDs) in series, with loop areas randomly distributed between 6 $\mu$m$^{2}$ and 60 $\mu$m$^{2}$. The DC transfer factor is $\sim450\ VT^{-1}$ at optimal bias and temperature, and the maximum voltage swing $\sim2.5\ mV$. We show that such a SQIF detects RF signals up to 150 MHz. It presents linear characteristics for RF power spanning more than five decades, and non-linearities develop beyond $P_{RF}=-35\ dBm$ in our set-up configuration. Second-harmonic generation has been shown to be minimum at the functioning point in the whole range of frequencies. A model has been developed which captures the essential features of the SQIF RF response.

\end{abstract}

\maketitle

\section{Introduction}

To develop the next generation  of  analog RF front end devices in wireless communications and radars, ultra wide bandwidth, compactness, high linearity, high sensitivity, and large dynamic range are essentials parameters. Those characteristics are particularly difficult to satisfy simultaneously in conventional antennas. Indeed, such devices use a resonance to amplify the voltage induced at the antenna terminals by the incident electromagnetic wave. The resonance is usually obtained either by a specific geometrical configuration and/or by the development of a complex tuning network leading in both cases to a strong reduction of the operation bandwidth. One way to overcome the limitations of classical antennas is the use of frequency independent sensitive devices such as Superconducting QUantum Interference Devices (SQUID), a superconducting loop interrupted by two Josephson Junctions (JJ)\cite{Clarke:2005tz}. They  are highly sensitive to an applied magnetic flux over a very large bandwidth, from DC to tens of GHz depending on the technology used for the fabrication of the JJ. Moreover, since the magnetic field of the wave is detected, and not the electric one as in conventional RF detector, SQUIDs can have sub-wavelength sizes, and therefore be very compact, keeping high sensitivity.

However the use of a single SQUID to perform efficient RF detection is limited by the low voltage swing across the interferometer, leading to a reduced range of linearity and rather low transfer factor $\partial V/\partial B$ (V is the voltage and B the applied magnetic field). The latter can be enhanced either by a larger SQUID loop area or by the adjunction of a surrounding flux transformer or flux concentrator, at the expense of the dynamic range. Moreover, because of the periodic response of the SQUID to the magnetic flux, a feed-back loop is mandatory to operate the device around a fixed functioning point and avoid flux jumps, which severely limits the bandwidth of the system as well\cite{Clarke:2005tz}.

In order to overcome those limitations, series arrays of $N$ SQUIDs have been proposed as RF amplifiers\cite{Welty:1991vk,Drung:2005hd,Cybart:2014et,Kornev:2017kra}. Both the transfer factor and the voltage swing increase as a function of $N$ while the output voltage noise increases as $\sqrt{N}$, which is favorable. For a large number of SQUIDs, it becomes possible to avoid the use of  a feedback electronic. In arrays of identical SQUIDs the periodicity of  the magnetic voltage response remains present.  Arrays known as Superconducting Quantum Interference Filters (SQIFs) were proposed in this context, to perform an absolute measure of the magnetic field \cite{Carelli:1997kv}.

Since the pioneer work of Oppenlander \textit{et al}\cite{Oppenlander:2000hz,Oppenlander:2003fc},  SQIFs appear as promising devices for cryogenic electronics. A SQIF is an array of SQUIDs  with loops of different sizes such as its response to an applied magnetic field is non-$\Phi_{0}$ periodic, where $\Phi_{0}=h/2e$ is the flux quantum. It belongs to the growing family of superconducting devices based on multiple Josephson Junctions arrays (see for a review Cybart \textit{et al}\cite{Cybart:2017bx} and references therein). The interest of SQIFs as compared to regular SQUIDs arrays lies in the single valued response to external magnetic field. It makes possible the realisation of highly sensitive absolute value magneto-sensors\cite{Caputo:2005cq,Schultze:2003fb,Oppenlander:2003hp,Schultze:2003vl}, and opens the route to high frequency compact detectors and Low Noise Amplifiers (LNA). The bandwidth of a SQIF is not limited by the feed-back loop, but in principle by the gap of the superconductors which is an upper bound (up to a few THz for some materials), and by the electrodynamics of the JJ and the superconducting circuit (more in the hundreds of GHz range). As a consequence, SQIFs are very good candidates to detect and amplify RF waves. 

In the recent years, very promising results have been reported in the literature about high-frequency devices based on SQIFs. Using Nb based technology, Kornev \textit{et al} made pulse amplifiers or drivers in the context of Rapid Single Flux Quantum (RSFQ) logic working at 100 MHz \cite{Kornev:2007bm}, and proposed advanced SQIFs architectures for microwave applications in general\cite{Kornev:2009fj}. Wide-band microwave LNA with a power gain of 20 dB from 8 to 11 GHz\cite{Prokopenko:2013to}, antennas in the near field\cite{Prokopenko:2015dx} at 9 GHz and active electrically antennas\cite{Kornev:2017hva} are being developed. 

These devices work at liquid helium temperature. This limits their applications, and High Tc Superconductors (HTS) appear as good candidates to make RF devices with SQIFs. On the one hand the operating temperature being higher, the cryogenics is simpler and cheaper. On the other hand their superconducting gap is an order of magnitude larger than the one of Low Tc  (LTS) materials, and so is the cut-off frequency of devices. Schultze \textit{et al} successfully made the first HTS SQIF\cite{Schultze:2003fb} that can be operated using commercial miniature cryocoolers\cite{Oppenlaender:2005fk}. In a first series of experiments, the JJ used in HTS SQIFs were fabricated using the so called Grain Boundary Junctions (GBJ) technology. Kornev \textit{et al}\cite{Kornev:2007bm} and  Kalabukhov \textit{et al}\cite{Kalabukhov:2008id} reported on a SQIF amplifier working at 100 MHz, while Shadrin \textit{et al}\cite{Shadrin:2008kka} estimated a 20 dB power gain at 1-2 GHz. Caputo \textit{et al} showed quadratic mixing using SQIFs up to 20 GHz\cite{Caputo:2007cs,Caputo:2006jg}. For practical reasons, it is difficult to make SQIFs with more than a few hundreds loops with the GBJ technology. Step-edge technology to fabricate HTS JJ is a most scalable process. Mitchell \textit{et al} succeeded in producing high $I_{c}R_{n}$ product JJ ($I_{c}$ is the critical current and $R_{n}$ is the resistance) with this technology\cite{Mitchell:2010el}. The $I_{c}R_{n}$ product sets the maximum operating frequency of the device through the Josephson relation $f_{J}=I_{c}R_{n}/\Phi_{0}$. They recently operated a 20 000 JJ SQIF with a transfer factor $\partial V_{DC}/\partial B \sim 1500\ V/T$ up to 30 MHz\cite{Mitchell:2016in}. 

An alternative technology to make HTS SQUIDs arrays with a large number of JJ is the irradiation technology\cite{Katz:1998gh,Bergeal:2005jna}, developed both for DC and high frequency operation\cite{Malnou:2012gt,Malnou:2014cp,Sharafiev:2018hc}. Large arrays have been produced\cite{Cybart:2008ff,Cybart:2009fc,Cybart:2014et} with up to 36 000 JJ. We recently showed that a 4000 JJ SQIF can have a transfer factor $\partial V_{DC}/\partial B \sim 1000\ V/T$\cite{Ouanani:2014cu,Ouanani:2016cr} which is very encouraging. We present here the behaviour of such arrays at frequencies up to 200 MHz.

\section{Experimental results}

\begin{figure}[h]
\includegraphics[scale=0.5]{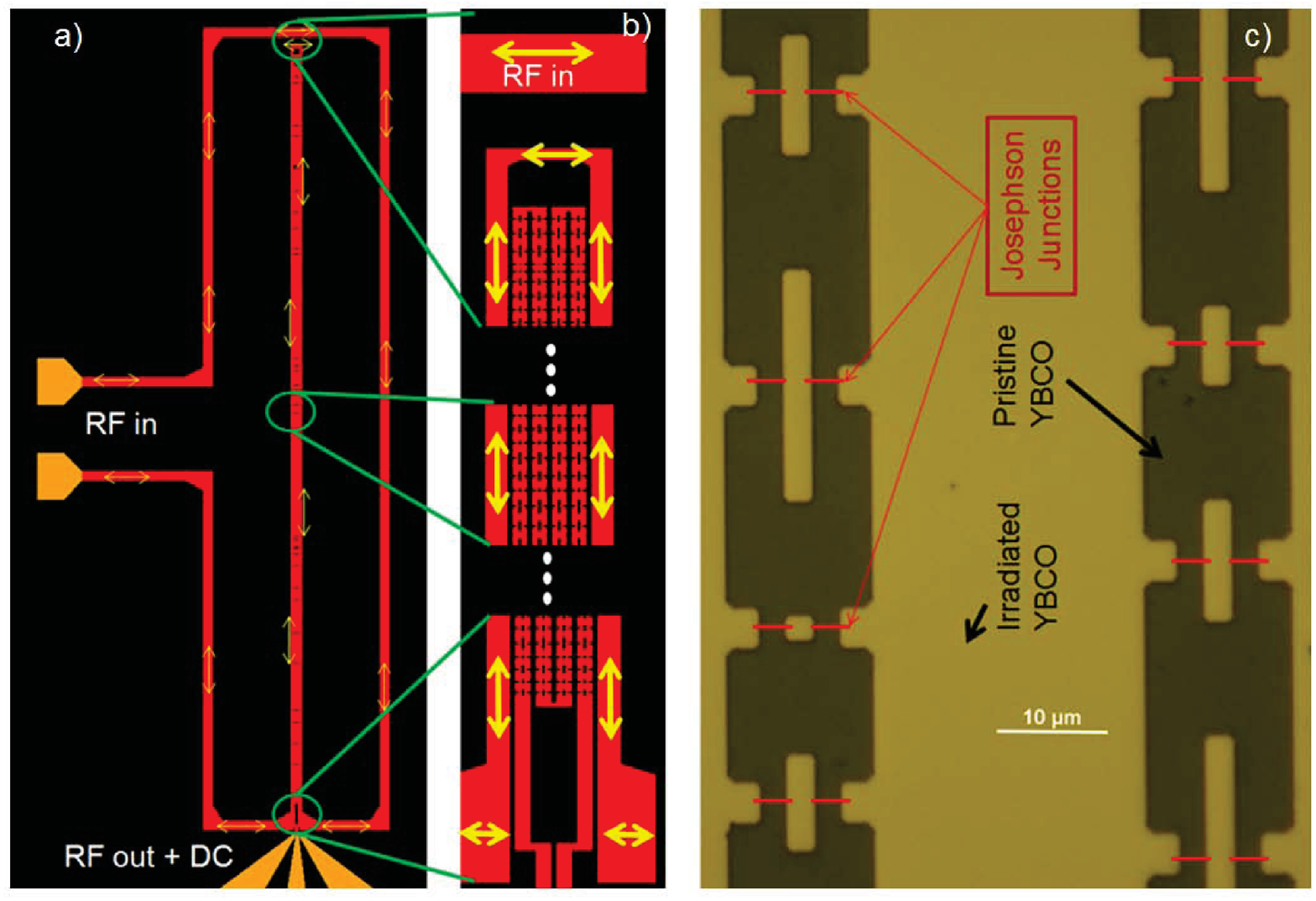} 
\caption{
\textit{(a)} Sketch of the device. A 1000 SQUIDs series SQIF is inserted within a RF loop. RF signal is applied via the loop (gold contacts on the left), and detected by the SQIF (gold contacts at the bottom connected to a CPW line). \textit{(b)} Zoom on the central part of the device. The SQIF is folded into a meander line, and the input loop is placed along it. Yellow arrows symbolise the RF excitation current. \textit{(c)} Optical picture of the central part of the SQIF which shows 8 individual SQUIDs. Irradiated YBCO zones (light color) are insulating after the first high dose irradiation. In a second step, JJ will be formed by low dose irradiation at the place shown by red lines.
}
\label{Figure1}
\end{figure}

We designed and fabricated HTS SQIFs for DC and RF measurements using the ion irradiation technique. Detail of the fabrication are described in previous papers\cite{Bergeal:2005jna,Bergeal:2006dd,Bergeal:2007jc,Ouanani:2014cu,Ouanani:2016cr} and summarised here. We start with a commercial\footnote[1]{Ceraco gmbh.} 150 nm thick c-axis oriented \YBCO (YBCO) film on a sapphire substrate covered by an in-situ 100 nm thick gold layer to insure ohmic contacts. After removing the gold layer by Ar-ion beam etching except on the contact pads, a photoresist is deposited on top of the YBCO layer to protect it from the subsequent ion-irradiation, and patterned to define the superconducting parts of the SQIF and the RF loop (see Figure \ref{Figure1} \textit{(a)}). 110 keV oxygen-ion irradiation at a dose of $5\times10^{15}\ $ions/cm$^{2}$ is then performed which makes the unprotected part insulating. A PMMA (polymethylmethacrylate) resist is then deposited all over the sample, and narrow (40 nm wide) slits are opened across each arm of the superconducting loops (individual SQUIDs) by electron beam lithography. A second 110 keV oxygen ion irradiation performed at lower dose ($3\times10^{13}\ $ions/cm$^{2}$) defines the  JJ (see Figure \ref{Figure1} \textit{c}).
This fabrication technique is very flexible, scalable, and allows the realisation of large and complex structures. The one which is presented here is a 1000 SQUIDs series array SQIF, with loop areas ranging from 6 to 60 $\mu$$m^{2}$ with a pseudo-random distribution, folded in a meander line, which is surrounded by a RF line (Figure \ref{Figure1}) to induce AC current in the system. The width of the SQUIDs arms is 2 $\mu$$m$ in the vicinity of the JJs.
To insure simultaneous DC and AC measurements, the sample is mounted on a Printed-Circuit Board (PCB) with Coplanar Wave Guides (CPW) lines for input and output RF signals, and DC pads. Short wire bonding between the sample and the PCB are then made, and the whole system is placed in a cryogen-free cryostat with no specific magnetic shielding. The external magnetic field is produced by Helmoltz coils mounted close to the sample. The RF signal is isolated from the DC part by a surface mounted bias-tee, pre-amplified at low temperature (100 K), and measured with a spectrum analyser.

\begin{figure}[h]
\includegraphics[scale=0.5]{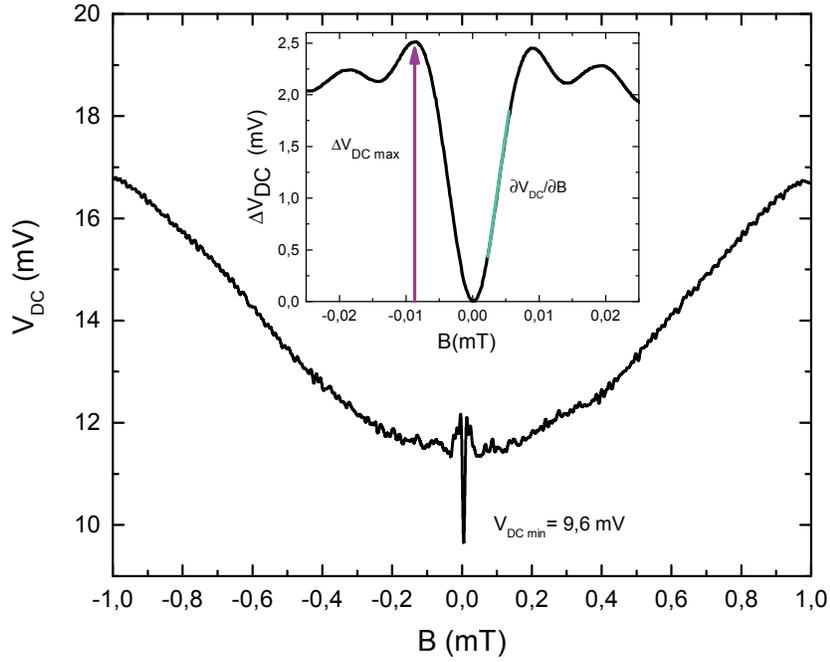} 
\caption{
DC characterisation of the 1000 SQUIDs series SQIF. The output DC voltage $V_{DC}$ as a function of the applied magnetic field B, for optimal bias ($I=65\ \mu$$A$) and at optimal temperature ($T=67.5\ K$). The SQIF characteristic dip around B=0  is observed. The minimum value $V_{min}$ is not at B=0 because of the offset field due to the unshielded environment. \textit{Inset} Black curve : normalised plot of the data at low magnetic field : $\Delta V_{DC}=V_{DC}-V_{min}$ as a function of B, where B is now the true field on the device after subtraction of the offset field. The maximum slope $V_{B}=\mid\partial\ V/\partial B\mid_{max}\sim450\ VT^{-1}$ (green line) and the maximum voltage swing $\Delta V_{DC max}\sim\ 2.5\ mV$ (purple arrow).}
\label{Figure2}
\end{figure}

We first focused on DC characteristics of the SQIF. As previously published, the Josephson regime in such device is observed below a coupling temperature $T_{J}\sim73\ K$\cite{Ouanani:2016cr}. A SQIF behaviour is observed under current bias larger than the critical current $I_{c}$ : the voltage $V_{DC}$ across the device presents a pronounced minimum around zero magnetic field  (Figure \ref{Figure2}). Due to the unshielded environment, the minimum value $V_{min}$ is not for $B=0$ exactly. The inset Figure \ref{Figure2} shows the voltage swing $\Delta V_{DC}=V_{DC}-V_{min}$ as a function of $B$ where the field offset has been subtracted. Hereinafter, this magnetic field offset has been systematically removed. The performance of the SQIF can be evaluated by the  voltage swing $\Delta V_{DC max}=max(\Delta V_{DC})$ and the maximum transfer factor $V_{B}=\mid\partial\ V/\partial B\mid_{max}$ at the inflexion point. Both of them are maximum for a given temperature and a given bias current\cite{Ouanani:2014cu,Ouanani:2016cr}. Figure \ref{Figure2} shows the $\Delta V_{DC}$ vs $B$ curve corresponding to these optimal values ($T=67.5\ K$ and $I=65\ \mu A$) : $\Delta V_{DC max}\sim\ 2.5\ mV$ and $V_{B}\sim\ 450 \ VT^{-1}$ are roughly half the values already reported for a 2000 SQUID SQIF made in the same conditions\cite{Ouanani:2014cu,Ouanani:2016cr} as expected, and compare favourably with previous reports for HTS devices\cite{Oppenlaender:2005fk,Mitchell:2016in}. 

We then measured the RF response of the sensor by superimposing a RF magnetic field and a slowly swept DC magnetic field $B$. A continuous wave (CW) input RF signal of frequency $f_{0}$ at a power level $P_{RF}$ is applied through the superconducting planar loop. The DC ($V_{DC}$) and the RF ($V_{RF}$) voltages are recorded simultaneously after amplification. In this study, the frequency $f_{0}$ ranges from 100 kHz to 200 MHz, and the input RF power $P_{RF}$ from -85 dBm to -10 dBm.

\begin{figure}[h]
\includegraphics[scale=0.5]{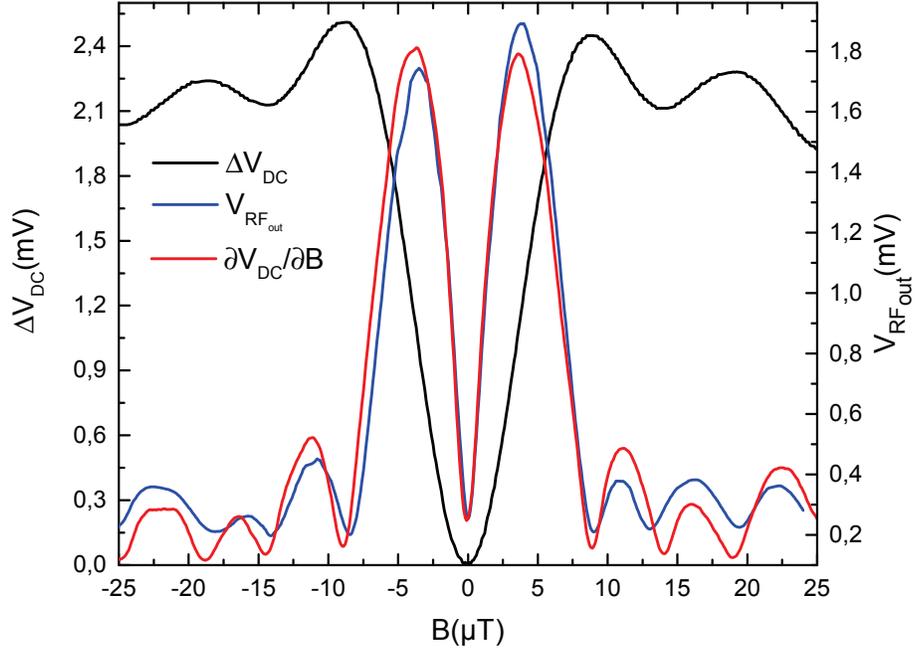}  
\caption{
Simultaneous measurement of the DC $\Delta V_{DC}$ (black curve) and RF $V_{RF}$ (blue curve) of a 1000 JJ series SQIF as a function of the magnetic field B. RF frequency is $f_{0}=30.02\ MHz$ and RF power is $P_{RF}=-50\ dBm$. Temperature (67.5 K) and bias current (65 $\mu$A) correspond to optimal conditions. In red is shown the derivative $\partial \Delta V_{DC}/\partial{B}$, which matches the RF response as expected. 
}
\label{Figure3}
\end{figure}

\begin{figure}[h]
\includegraphics[scale=0.5]{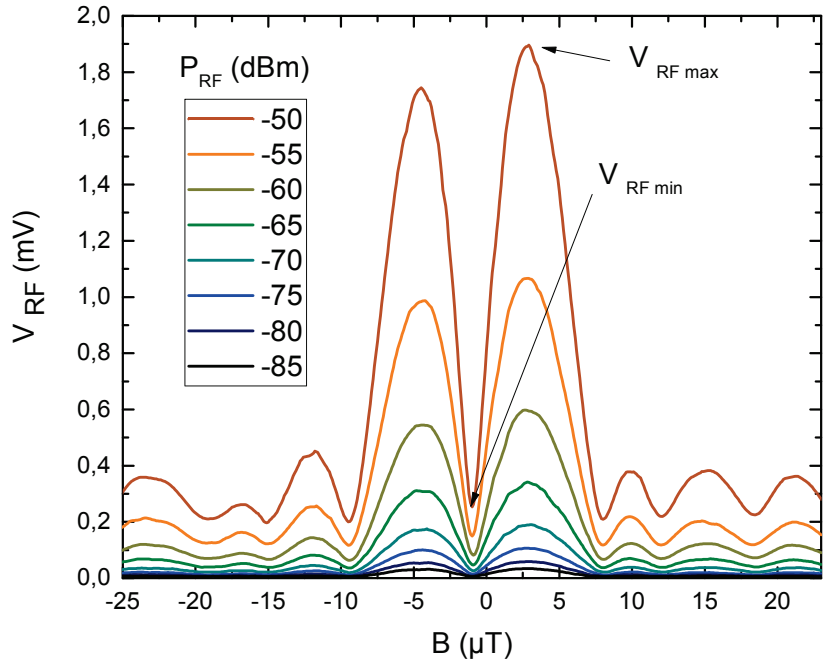} 
\caption{
RF signal $V_{RF}$ of a 1000 SQUIDs series SQIF as a function of the magnetic field B for different RF powers from - 85 dBm to - 50 dBm. Temperature (67.5 K) and bias current ($65\ \mu A$) are optimal. RF frequency is $f_{0}=30.02\  MHz$. The maximum $V_{RF max}$ and minimum $V_{RF min}$ values are indicated by arrows.
}
\label{Figure4}
\end{figure}

\begin{figure}[h]
\includegraphics[scale=0.4]{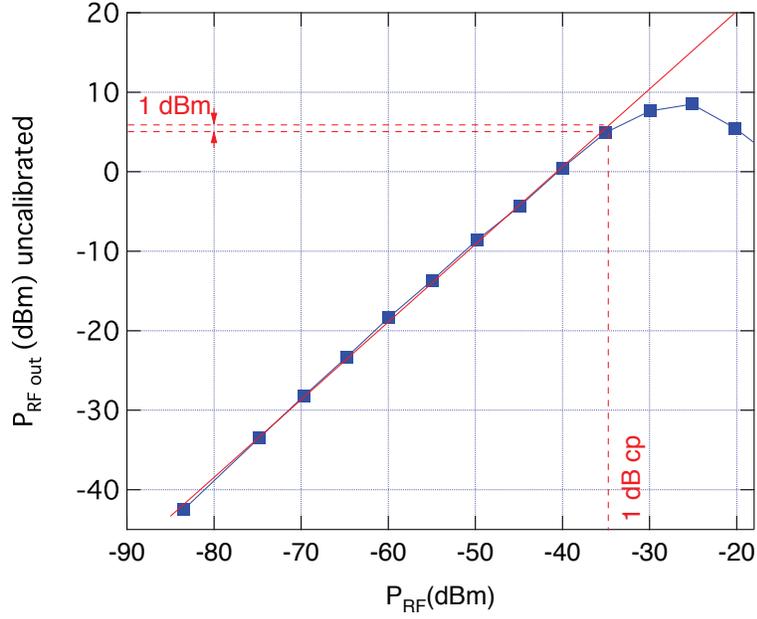}
\caption{
$P_{RF out}$ as a function of $P_{RF}$ for a 1000 SQUIDs series SQIF. Temperature (67.5 K) and bias current ($65\ \mu A$) are optimal.  $P_{RF out}$ is the square of the maximum RF voltage $V_{RF max}$, but uncalibrated. RF frequency is $f_{0}=10\  MHz$ and the RF power ranges from -85 dBm to -20\ dBm. A linear behaviour is observed over 5 decades in RF power (red line). The one dB compression point is shown, setting the limit of the linear regime to -35 dBm.
}
\label{Figure15}
\end{figure}

Figure \ref{Figure3} shows $V_{RF}$ (blue curve) and $\Delta V_{DC}$ (black curve) as a function of $B$ in the optimal conditions (temperature and bias current) for $f_{0}=30.02\ MHz$ and $P_{RF}=-50\ dBm$. The total magnetic field applied on the SQIF is 
$B_{tot}=B+b_{RF}\sin{(2\pi f_{0}t)}$, 
where $b_{RF}$ is the RF magnetic field amplitude proportional to the square root of $P_{RF}$. For small $b_{RF}$, one can make a first order Taylor expansion of the output signal, and $V_{RF}\propto \partial \Delta V_{DC}/\partial B$. As expected in a linear regime, the RF signal appears as the derivative of the DC one $\partial \Delta V_{DC}/\partial B$ which has been numerically calculated and shown in Figure \ref{Figure3} (red curve). 
The RF output signal increases with $P_{RF}$ while keeping the same pattern as a function of the DC magnetic field $B$, in a wide range of input power, as shown in Figure \ref{Figure4} for $P_{RF}$ in the range -85 dBm to -50 dBm. To characterise the linearity of the SQIF response, we focused on the output power $P_{RF out}\propto V_{RF}^2$  at the optimum operating point. Interestingly, as seen in the inset Figure \ref{Figure2}, this point corresponds both to the maximum of $V_{RF}$ (called hereinafter $V_{RF max}$) and the centre of the most linear part of the $\Delta V_{DC}$ vs $B$ curve. In Figure \ref{Figure15} is  shown $P_{RF out}\propto V_{RF max}^2$ (expressed in dBm, not calibrated) as a function of $P_{RF}$ for a frequency $f_{0}=10\ MHz$. The linear behaviour extends on more than five decades, up to $P_{RF}=-35\ dBm$ (one dB compression point criteria). This result is valid for the whole range of frequencies studied here.

\begin{figure}[h]
\includegraphics[scale=0.5]{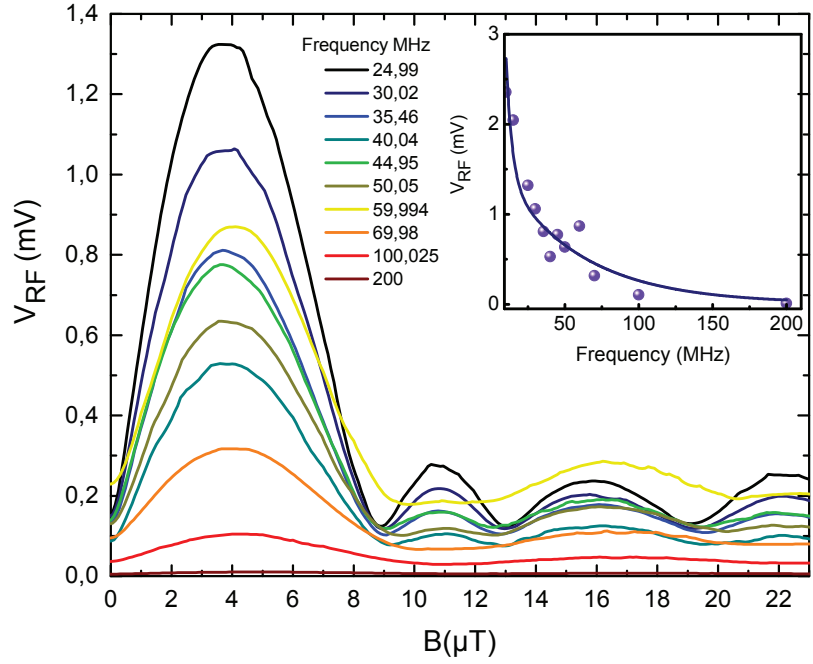} 
\caption{RF signal $V_{RF}$ of a 1000 SQUIDs series SQIF as a function of the magnetic field B for different excitation frequencies $f_{0}$ ranging from 25 MHz to 200 MHz at a constant input power $P_{RF}=-55\ dBm$. Temperature (67.5 K) and bias current ($65\ \mu A$) are optimal. \textit{(Inset)} Maximum value $V_{RF max}$ as a function of $f_{0}$. The solid line is a guide for the eyes.}
\label{Figure11}
\end{figure}

We increased the frequency up to $f_{0}=200\ MHz$ as shown in Figure \ref{Figure11}. For a constant RF input power $P_{RF}=-55\ dBm$, the amplitude of the output signal $V_{RF}$ decreases as $f_{0}$ is increased, and dies out between 100 and 200 MHz, as seen in the inset of Figure \ref{Figure11}. This limitation may be not intrinsic to the HTS SQIF, since the RF circuitry is not fully optimised in this series of experiments. This result clearly shows that HTS SQIFs made by ion irradiation can operate at least up to 150 MHz. The maximum operation frequency reported for a HTS SQIF made with step-edge JJ\cite{Mitchell:2016in} is 30 MHz, while GBJ based HTS SQIF can operate up to 100 MHz\cite{Kornev:2007bm,Kalabukhov:2008id}. 

\begin{figure}[h]
\includegraphics[scale=0.5]{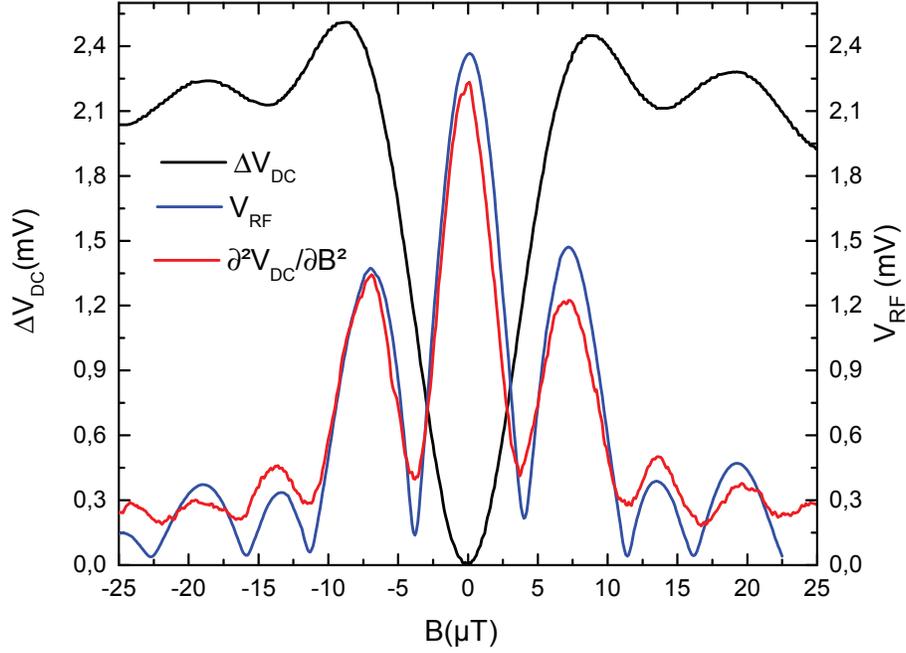} 
\caption{
Simultaneous measurement of the DC $\Delta V_{DC}$ (black curve) and RF $V_{RF}$ (blue curve) signals of a 1000 SQUIDs series SQIF as a function of the magnetic field B. The input RF frequency is $f_{0}=30.02\ MHz$ and the RF Power is $P_{RF}=-30\ dBm$. Temperature (67.5 K) and bias current ($65\ \mu A$) correspond to optimal conditions. RF is detected at $2\times f_{0}=60.04\ MHz$. In red is shown the second derivative $\partial^{2} \Delta V_{DC}/\partial{B^{2}}$, which matches the RF response at $2\times f_{0}$ as expected. 
}
\label{Figure5}
\end{figure}

\begin{figure}[h]
\includegraphics[scale=0.5]{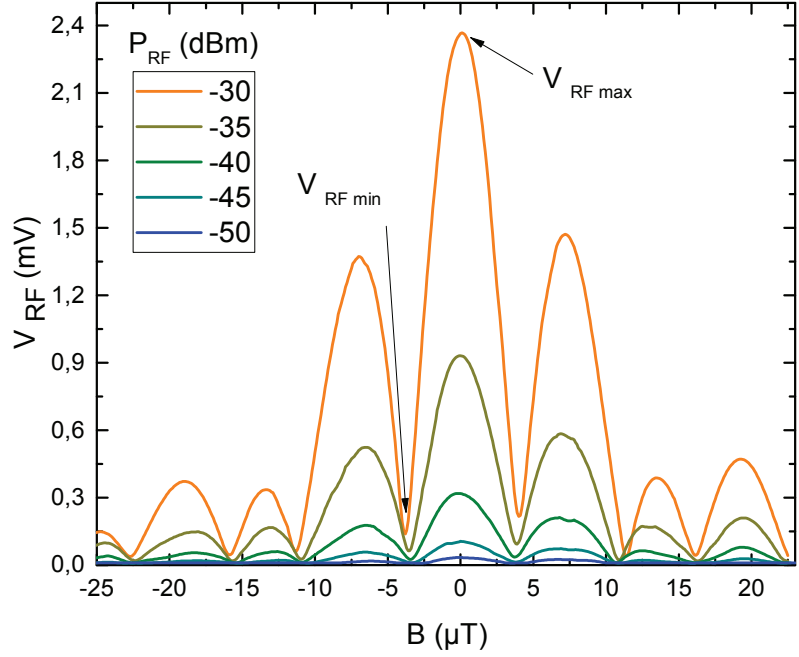}  
\caption{
RF signal $V_{RF}$ of a 1000 SQUIDs series SQIF as a function of the magnetic field B for different RF powers $P_{RF}$ ranging from - 50 dBm to - 30 dBm. Temperature (67.5 K) and bias current ($65\ \mu A$) are optimal. RF excitation frequency is $f_{0}=30.02\ MHz$. The signal is detected at  $2\times f_{0}=60.04\ MHz$. The maximum $V_{RF max}$ and minimum $V_{RF min}$ values are indicated by arrows.
}
\label{Figure6}
\end{figure}

Since the $\Delta V_{DC}$ vs $B$ curve is essentially non-linear, it generates harmonics when the SQIF is submitted to an AC radiation. This is indeed observed as shown in Figure \ref{Figure5}. It is important to evaluate its contribution to the output signal. An input signal at $f_{0}=30.02\ MHz$ is sent and detected at twice the frequency $2\times f_{0}=60.04\ MHz$ (blue line), and compared with $\partial^{2} \Delta V_{DC}/\partial B^{2}$ (red line). The agreement between the two is good.This behaviour is observed in the whole frequency range studied here. The frequency doubling is observed on a wide range of input power below typically $P_{RF}\sim -35\ dBm$ (see Figure \ref{Figure6}). Beyond this threshold, departure from quadratic operation emerges, and the RF output voltage $V_{RF}$ does not reflect the input one. It is worth noticing that at the operating point, \textit{i.e.} when the transfer factor $V_{B}=\mid\partial\ V/\partial B\mid_{max}$ is maximum, the second harmonic signal is always minimum in all conditions.

\begin{figure}[h]
\includegraphics[scale=1]{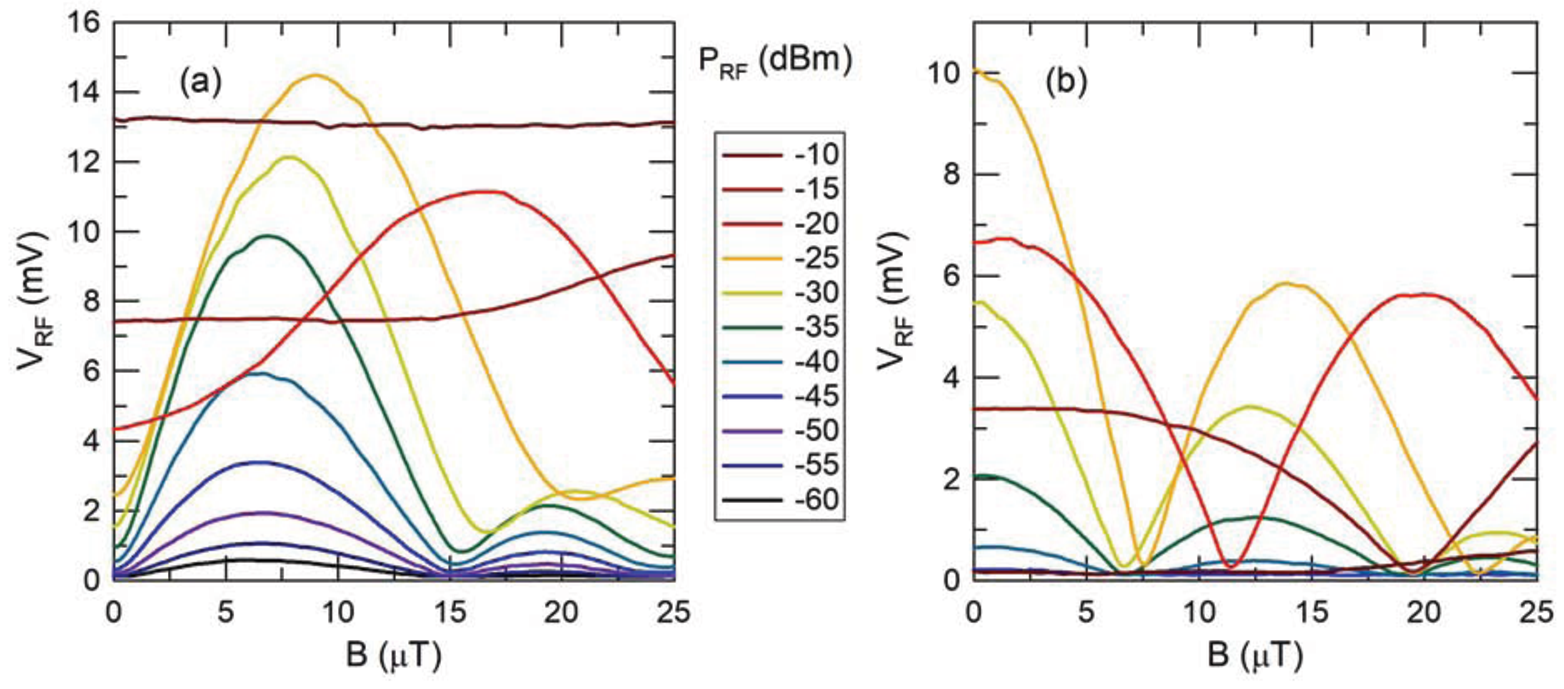} 
\caption{
\textit{(a)} RF signal $V_{RF}$ of a 1000 SQUIDs series SQIF as a function of the magnetic field $B\geq0$ for increasing RF power $P_{RF}$, at temperature $T=67.2\ K$ and bias current $70\ \mu A$. The frequency is $f_{0}=101\ kHz$. A linear regime is observed for $P_{RF}$ ranging from -60 to -35 dBm, while a non-linear regime is evidenced beyond - 35 dBm. \textit{(b)} The same behaviour is observed for the second harmonic signal.
}
\label{Figure7}
\end{figure}

We studied in more detail the development of the non-linearity. Figure \ref{Figure7} \textit{(a)} shows $V_{RF out}$ as a function of the applied DC magnetic field $B$ at temperature $T=67.2\ K$ and bias current $I=70$ $\mu$$A$, for an AC frequency $f_{0}=101\ kHz$ and  power $P_{RF}$ ranging from -60 to -10 dBm. At low power, typically for $P_{RF}\leq -35\ dBm$, a linear regime is observed : the pattern of the curve is kept constant as $P_{RF}$ is increased, and its amplitude grows with the RF power. This can be seen by following the maximum of the curve $V_{RF max}$ for instance. Beyond $-35\ dBm$ the curve departs from the original pattern, $V_{RF max}$ starts saturating and then decreases as $P_{RF}$ is increased and eventually, all structures disappear when reaching -10 dBm : this is the non-linear regime. The two regimes are observed in the second harmonic signal as well. In Figure \ref{Figure7} \textit{(b)}, the RF signal is plotted vs $B$ for an excitation signal at $f_{0}=101\ kHz$ detected at $2\times f_{0}=202\ kHz$ at different RF powers $P_{RF}$. The linear regime is seen below  $-35\ dBm$. The transition from linear to non-linear regime is observed for all the frequencies studied here, and the threshold is found to be around $P_{RF}$=30-35 dBm.

\section{Modelling}

We performed numerical simulations in order to describe in more detail the behaviour of the SQIF, in particular the transition from linear to non linear regime. As an example, we show here the calculations made in the case of an RF signal at frequency $f_{0}=101\ kHz$ (as in the data shown above), but they can be extended to all frequencies that we used. The idea is simply to add the RF and DC magnetic fields seen by the SQIF. As stated above the total field will be $B_{tot}=B+b_{RF}\sin{(2\pi f_{0}t)}$, where $B$ is the DC magnetic field and $b_{RF}$ the amplitude of the RF field. We can estimate it by calculating the magnetic field produced by the RF line on the closest SQUIDs of the array : 
\[b_{RF}\ =\left(\frac{\mu_{0}}{2\pi r}\right)\cdot I_{RF}\]
where $r$ is roughly 20 $\mu$m (see Figure \ref{Figure1}) and $I_{RF}$ is the RF current. For a circuit with an impedance matched to $Z=50\  \Omega$, the RF power is $P_{RF}=ZI_{RF}^{2}$. Therefore, $b_{RF}$ relates to $P_{RF}$ via $b_{RF}\sim 1.5\cdot 10^{-3} \sqrt{P_{RF}}$ in \textit{SI} units.

We now calculate the RF signal $V_{RF}$ measured across the SQIF.  We start with the DC curve $\Delta V_{DC}$ vs $B$ recorded simultaneously with the AC signal and presented in the inset of Figure \ref{Figure14} \textit{(b)} for positive magnetic field only. A function $G(B)$ is chosen to fit it accurately (see red solid line in Figure \ref{Figure14} \textit{(b)}) : 

\[\Delta V_{DC}= G(B)=w_{0}(1-\frac{\sin{(w_{1}B)}}{w_{1}B})-(w_{2}B^2+w_{3}B+w_{4})\exp{(-w_{5}B^2})\]

where $w_{i}$ are fitting coefficients\footnote[2]{$w_{0}=1.68\cdot10^{-3}$ ; $w_{1}=317.2$ ; $w_{2}=-8.545$ ; $w_{3}=-6.11\cdot10^{-3}$ ; $w_{4}=-1.20\cdot10^{-5}$ ; $w_{5}=3102$}

Upon RF irradiation, we expect the total voltage across the SQIF to be $V_{tot}=G(B_{tot})$. The amplitude of the Fourier components of $V_{tot}$ at $f_{0}$ and $2\times f_{0}$ give the RF output amplitude $V_{RF}$ and the second harmonic.

\begin{figure}[h]
\includegraphics[scale=1]{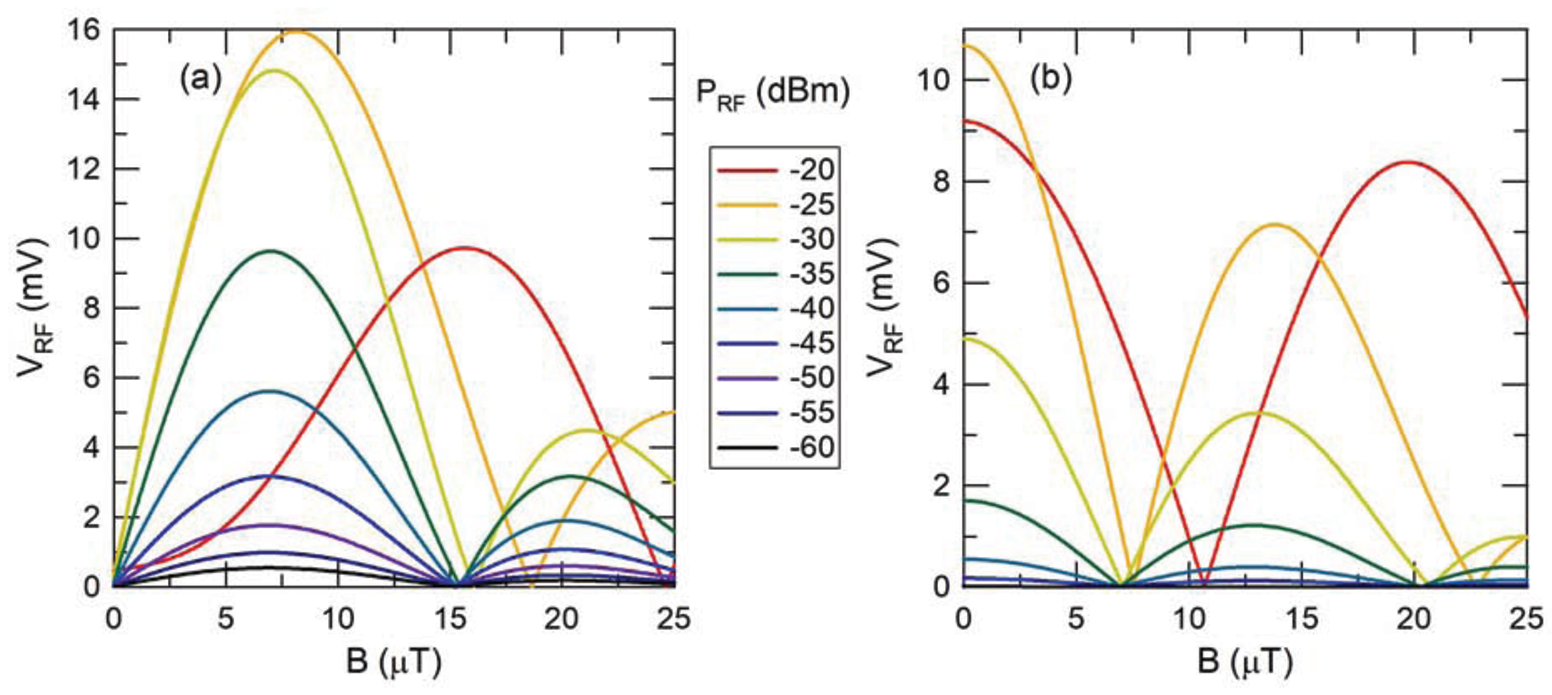}
\caption{
Simulation of the RF signal $V_{RF}$ as a function of the magnetic field B, for different RF power $P_{RF}$ ranging from -60 to -20 dBm (see text for detail). \textit{(a)} Main component. \textit{(b)} Second harmonic.
}
\label{Figure12}
\end{figure}

\begin{figure}[h]
\includegraphics[scale=0.7]{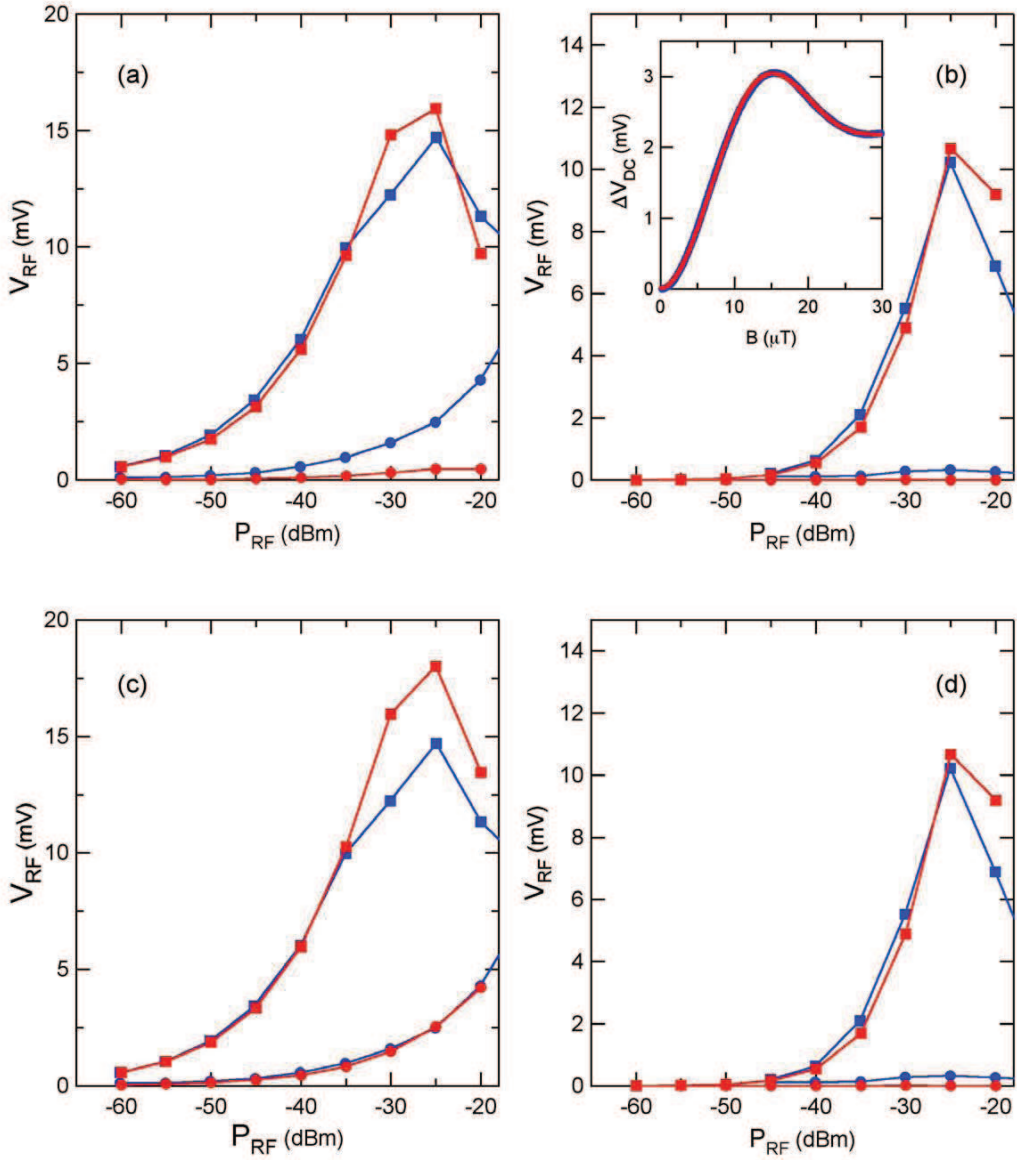}
\caption{
\textit{(a)} Maximum $V_{RF max}$ (solid squares) and minimum $V_{RF min}$ (solid circles)  of the RF voltage as a function of the RF power $P_{RF}$. Data measured at $f_{0}= 101\ kHz$ are shown in blue, and simulations in red. \textit{(b)} Maximum (solid square) and minimum (solid circle) of the second harmonic RF voltage as a function of the power $P_{RF}$. Data are shown in blue, and simulation in red. \textit{(Inset)} DC response of the SQIF ($\Delta V_{DC}$ vs magnetic field B) measured while recording the AC curve with $f_{0}= 101\ kHz$ at temperature $T=67.2\ K$ and bias current $I=70\ \mu A$ : in blue the data and in red the fit function $G(B)$ used for the simulation. \textit{(c)} Same data as in \textit{(a)}. Simulation now includes the direct coupling (see text). \textit{(d)} Same data as in \textit{(b)}. Simulation now includes the direct coupling (see text).
}
\label{Figure14}
\end{figure}

The result of the calculation is shown in Figure \ref{Figure12}, and compared to the measurements (Figure \ref{Figure7}). The main features are reproduced, such as the linear behaviour at low power and the non monotonic behaviour of $V_{RF max}$. To make a quantitative comparison, we introduced a single gain parameter as followed : for the lowest RF power, namely $-60\ dBm$, we made the maxima $V_{RF max}$ coincide numerically. We then calculated the whole set of curves presented in Figure \ref{Figure12} \textit{(a)}. The agreement with the measured data for different RF powers is very good. This simple model shows that the non linearity originates in the amplitude of the magnetic RF field becoming comparable to the DC one for $P_{RF}\sim -35\ dBm$. Under the same assumptions, we computed the second harmonic signal, and plotted it in Figure \ref{Figure12} \textit{(b)}. The agreement with the data of Figure \ref{Figure7} is also striking.

In Figure \ref{Figure14}, we plotted $V_{RF max}$ (solid squares) and the amplitude at null DC magnetic field $V_{RF min}$ (solid circles), as a function $P_{RF}$ for the first \textit{(a)} and the second \textit{(b)} harmonic signals. In blue are the data and in red the computed values. The agreement between the calculation and the data is good for $V_{RF max}$, but rather poor for $V_{RF min}$ : the latter increases significantly with $P_{RF}$ in the data, while staying almost zero in the calculation. If we focus on the second harmonic signal depicted in Figure 
\ref{Figure14} \textit{(b)}, we observe that the agreement is good both for the maximum and the minimum of the signal. Put together, this means that a significant direct inductive coupling occurs between the input RF line and the detection line, by-passing the SQIF itself. It is linear, and therefore not seen in the second harmonic signal. 

We tested this hypothesis and introduced a direct coupling. The total RF voltage is now therefore : $V_{tot}=G(B_{tot})+C\cdot b_{RF}\sin{(2\pi f_{0}t)}$, where $C$ is the coupling constant that we set by matching the experimental and calculated $V_{min}$ values at $P_{RF}=-20\ dBm$ ($C = 2\cdot 10^{-2}$). The result is shown in Figure \ref{Figure14} \textit{(c)} and \textit{(d)}, for the first and second harmonic signals, respectively. The agreement between data and computation is much better, confirming our hypothesis.

\section{Conclusion}

We measured the static and radio-frequency magnetic response of a HTS SQIF with 1000 SQUIDs in series, fabricated by the ion irradiation technique. A transfer factor of $\partial \Delta V_{DC}/\partial B\sim450\ VT^{-1}$ was achieved in DC, with a maximum swing voltage of $\sim\ 2.5\ mV$. Coupled to a RF loop, the device presents an AC response up to 150 MHz in the experimental set-up that we used. Second-harmonic generation has been shown to be minimum at the functioning point in the whole range of frequencies. At low RF power, over five decades below $\sim\ -35\ dBm$, the device operates in a linear regime, and the output signal is proportional to the input one. Beyond this value, a non-linear regime is observed, where the modulation of the AC voltage with the applied magnetic field departs from the original one. A model has been developed which is in quantitative agreement with most of the data. By analysing the differences, we showed that direct coupling to the output RF line in our set-up measurement limits the performances of the SQIF device.
These results compare favourably with the ones presented in the literature for HTS SQIFs made by other methods. This shows that the ion irradiation technique provides an interesting route to make competitive HTS RF devices on a large scale.

\section{Aknowledgements}

The authors thank Yann Legall (ICUBE laboratory, Strasbourg) for ion irradiations. his work has been supported by ANRT and Thales through a CIFRE PhD fellowship n¡2015/1076, the T-SUN ANR ASTRID program (ANR-13-ASTR-0025-01), the Emergence Program from Ville de Paris and by the R\'egion Ile-de-France in the framework of the DIM Nano-K and Sesame programs.


\end{document}